\documentclass{elsart}

\usepackage[T1]{fontenc}
\usepackage[latin1]{inputenc}
\usepackage{epsfig}
\usepackage{graphicx}
\usepackage{verbatim}
\usepackage{amsmath}
\usepackage{comment}

\begin{document} 
\begin{frontmatter} 

\newcommand{\bd}[1]{ \mbox{\boldmath $#1$}  }
\newcommand{\xslash}[1]{\overlay{#1}{/}}
\newcommand{\sla}[1]{\xslash{#1}}
\newcommand{\qn}{\mathbf{q}}
\newcommand{\rn}{\mathbf{r}}
\newcommand{\Rn}{\mathbf{R}}
\newcommand{\elab}{$E_\mathrm{lab}$}

\title{$\alpha$-particle production in the scattering of $^6$He by $^{208}$Pb at energies around 
the Coulomb barrier}


\author[csic]{D. Escrig},
\author[huelva,lln]{A.M. S\'anchez-Ben\'itez},
\author[famn]{A.~M.~Moro\corauthref{cor2}}, 
\corauth[cor2]{Corresponding author}
\ead{moro@us.es}
\author[famn]{M.A.G. \'Alvarez},
\author[famn]{M.V. Andr\'es},
\author[lln]{C.~Angulo},
\author[csic]{M. J. G. Borge},
\author[lln]{J.~Cabrera},
\author[catania]{S.~Cherubini},
\author[lln]{P. Demaret},
\author[famn]{J.M.~Espino},
\author[catania]{P.~Figuera},
\author[england]{M.~Freer}, 
\author[huelva]{J.E.~Garc\'ia-Ramos}, 
\author[famn]{J.~G\'omez-Camacho},
\author[catania]{M.~Gulino},
\author[iran]{O.R.~Kakuee},
\author[huelva]{I.~Martel},
\author[england]{C. Metelko},
\author[huelva]{F. P\'erez-Bernal}, 
\author[iran]{J.~Rahighi},
\author[huelva,warsaw]{K. Rusek}, 
\author[leuven]{D.~Smirnov},
\author[csic]{O.~Tengblad},
\author[england]{V.~Ziman}

\address[csic]{Instituto de Estructura de la Materia, CSIC, E-28006, Madrid, Spain}
\address[huelva]{Departamento de F\'{\i}sica Aplicada, Universidad de Huelva, 
E-21071, Spain}
\address[lln]{Institut de Physique Nucl\'eaire and Centre de Recherches du Cyclotron, 
Universit\'e catholique de Louvain, B-1348 Louvain-la-Neuve, Belgium}
\address[famn]{Departamento de F\'{\i}sica At\'omica, Molecular y Nuclear, 
 Universidad de Sevilla,  Apdo. 1065, E-41080 Sevilla, Spain} 
\address[catania]{INFN Laboratori Nazionali del Sud, I-95123 Catania, Italy}
\address[england]{School of Physics and Astronomy, 
University of Birmingham, B15 2TT Birmingham, United Kingdom}
\address[iran]{Van de Graaff Laboratory, Nuclear Research Centre, 
AEOI, PO Box 14155-1339, Tehran, Iran}
\address[warsaw]{Department of Nuclear Reactions, 
Andrzej Soltan Institute for Nuclear Studies, Hoza 69, PL-00681 Warsaw, 
Poland}
\address[leuven]{Instituut voor Kern-en Stralingsfysica, University of Leuven, B-3001 Leuven, Belgium.}









\begin{abstract}
New experimental data from the scattering  of \nuc{6}{He}+\nuc{208}{Pb} at 
energies around and below the Coulomb barrier are presented. The 
yield of breakup products coming from projectile fragmentation  
is dominated by a strong group of $\alpha$ particles.  
The  energy 
and angular distribution of this group have been analyzed  
and compared with theoretical calculations. This analysis indicates 
that the $\alpha$ particles emitted at backward angles in this reaction 
are mainly due to two-neutron transfer to weakly bound states 
of the final nucleus. 

\end{abstract}

\end{frontmatter}

\section{Introduction \label{section:intro}}
Reactions involving  nuclei far from  stability  at energies around 
the Coulomb barrier
provide an excellent tool to study the novel properties of  exotic systems.  
The increase in intensity of radioactive ion beams  achieved along the past two decades has
contributed decisively to the improvement of  the accuracy of these experiments.
One of the exotic nuclei that has received more attention in recent years
is the \nuc{6}{He}. Its interesting Borromean structure, consisting of an 
$\alpha$ core plus two-weakly bound halo neutrons,  and its relatively long 
half-life (807 ms) makes  this nucleus  an excellent candidate for this kind
of experiments. Reactions induced by  
\nuc{6}{He} on several targets \cite{Agu00,DeY05,DiPiet04,Nav04,Raa04,Lian05} at energies around the 
Coulomb barrier  exhibit
some common features, such as a remarkably large cross section for the production 
of $\alpha$ particles. This effect is clearly associated with 
the weak binding of the halo neutrons, 
that favors the dissociation of the \nuc{6}{He} projectile in the nuclear 
and Coulomb field of the target.

To place our work in the appropriate context, we first review some recent experiments with \nuc{6}{He}
at Coulomb barrier energies. 
In the work of Aguilera {\em et al.}  \cite{Agu00} a simultaneous 
analysis of the elastic and two-neutron removal channels for the \nuc{6}{He}+\nuc{209}{Bi} 
reaction
revealed that, at energies below the Coulomb barrier, the reaction cross section is almost
exhausted  by the $\alpha$ channel while the complete fusion 
cross section is very small. In a more recent measurement of the same reaction \cite{DeY05}
in which  neutron-$\alpha$ coincidences were  recorded, the authors concluded that more than
half of the  $\alpha$ particles produced beyond the grazing angle arise from two-neutron transfer
to unbound states of the \nuc{211}{Bi} residual nucleus.  
Similar conclusions were achieved in an experiment  done by
Di Pietro {\em et al.}  \cite{DiPiet04}, where about 80\% of the measured $\alpha$ 
particles coming out from the
reaction  \nuc{6}{He}+\nuc{64}{Zn} were identified as coming from transfer or breakup.
As a further example, we mention the measurements of Navin {\em et al.}  for \nuc{6}{He}+\nuc{65}{Cu}
at $E_\mathrm{lab}=19.5$ and 30 MeV \cite{Nav04}. The \nuc{66}{Cu} yield is 
largely underestimated by statistical model  calculations, suggesting that an important 
fraction of these products have an origin different from fusion evaporation. The observation of 
the characteristic  $\gamma$ rays from heavy products (such as \nuc{65}{Cu}) in coincidence 
with 
projectile-like particles, 
confirm that these other processes could be  1n and 2n transfer followed by evaporation. 

Recently, Raabe {\em et al.}   \cite{Raa04}, measured fission fragments  
for the reaction \nuc{6}{He}+\nuc{238}{U} at energies around the fusion barrier. Based upon
kinematical considerations, they concluded that the large  observed yield for fission below the barrier is 
entirely due to a direct process, the two-neutron transfer. 
Their conclusions were supported by calculations performed in the distorted wave Born 
approximation (DWBA)  for the 
two-neutron transfer to excited states of the  \nuc{238}{U} target. 

In this work, we present new data for the breakup of \nuc{6}{He} on \nuc{208}{Pb} at energies
around the Coulomb barrier, measured at the  
CYCLONE RNB facility at the Centre de Recherche du Cyclotron (CRC) of
the Université catholique de Louvain (UCL), Louvain-la-Neuve, Belgium. The 
elastic scattering data from the same reaction 
has been analysed and presented before \cite{angel05,Kak06,angel06}.
A comprehensive optical model analysis of these  data   
revealed the existence of a long-range absorption effect, which is a clear indication of the
presence of reaction mechanisms that remove flux from the elastic channel at distances
well beyond the strong absorption radius. The same effect has been also reported
in other reactions induced by 
\nuc{6}{He} \cite{Agu01} and by other weakly bound nuclei, 
such as \nuc{17}{F}+\nuc{208}Pb \cite{Rom04}. The inclusion of a dynamic polarization
potential (DPP) in the  phenomenological projectile-target interaction
showed that  part of this long-range absorption effect arises from the distortion 
produced in \nuc{6}{He} due to the intense dipole Coulomb interaction \cite{May94}. 
The phenomenological optical model required a very large imaginary diffuseness
in order to reproduce the elastic 
data even after
the inclusion of the DPP. Thus, the nuclear interaction 
contributes also to this long-range absorption.

Given the dominance of the \nuc{4}{He} channel in these low energy reactions,
it is plausible to suggest that the  mechanisms responsible for the production 
of these fragments are also responsible for the long-range absorption effect. Guided by this
motivation, 
in this work we present an analysis of the two-neutron removal channel measured in 
the same experiment. In 
this experiment, the energy and scattering angle of the $\alpha$ particles emitted 
at backward angles were recorded.
The purpose of this work is to understand the reaction mechanisms which are relevant in the collision of
$^6$He on $^{208}$Pb, by examining  the angular and energy distributions of the $\alpha$ particles produced
in the collision. To do this, for each beam energy, the angular distribution as well 
as the  angle-integrated energy distribution of the  $\alpha$ particles have 
been evaluated and compared with theoretical calculations.


The paper is organized as follows. In section 2 we describe the experimental 
setup and analysis method, and discuss general features of the measured observables. In 
section 3, we compare these observables with theoretical 
calculations performed with the transfer to the continuum  method,  direct breakup calculations using 
continuum discretized coupled-channels  calculations, 
and  neutron transfer  calculations in DWBA. 
In section 4 we discuss our results and compare them with 
previous works.  Finally, in section 5 we present the summary and outlook of this work.

\section{\label{sec:general}Experimental setup and analysis procedure}
The scattering of a $^6$He beam on a $^{208}$Pb target was studied 
at the radioactive beam facility of the 
CRC/UCL at Louvain-la-Neuve. The $^6$He beam was produced
by the $^7$Li(p,2p)$^6$He reaction in a LiF powder target with a graphite container.
The atomic beam was ionized in an ECR source, purified by magnetic separation
and reaccelerated in the CYCLONE110 cyclotron at the CRC/UCL.  This technique provides a highly pure 
beam, essentially free of contaminants. The only contaminant ever observed in this beam has been reported in the   
the work of Miljanic {\em et al.} \cite{Mil00} where \nuc{4}{He}\nuc{1}{H}$_2^{+}$ ions were observed 
as an impurity in a 
17 MeV \nuc{6}{He} beam with a ratio of intensities of 1:5400. For the angular range covered in the 
present experiment, the contribution of this impurity turned out to be negligible.   The 
$^6$He beam was produced at laboratory energies 
of 14, 16, 18 MeV within the first harmonic of the accelerator with an average intensity of
4$\cdot $10$^6$ ions per second and at 22 MeV at the lowest limit of the second harmonic
with an intensity of 1.5$\cdot $10$^5$ ions per second. A high intensity 
$^4$He beam at laboratory energy of 12 MeV was used for normalization
of the elastic cross section.

The elastic data obtained from this experiment are
published elsewhere~\cite{angel06}. 
We briefly review here the main features of the setup.
For a more detailed description
of the experimental setup we refer to previous publications~\cite{angel05,angel06}.
The size of the $^{4,6}$He beams were reduced by passing
the beam through a set of two collimators of 5 mm and 7 mm diameter 
with the latter at 400~mm from the target.
The targets consisted of self-supporting foils of enriched $^{208}$Pb (87\%)
mounted on a movable ladder with a thickness of 0.950 mg/cm$^2$
for the 12 MeV $^4$He and 14,16,18 MeV $^6$He beams and 2.080 mg/cm$^2$ for the 22 MeV
$^6$He beam.
The latter was used to compensate for the low intensity of the beam.

The reaction products were measured using four LEDA detectors
in the standard form and six LEDA detectors in the
LAMP configuration covering angles in the forward direction
from 5$^{\circ }$ to 65$^{\circ }$. For a detailed description of 
the performance and efficiency of
these detectors see~\cite{dav00}. 
In the backward direction, the most relevant for this analysis, the 
DINEX telescope array~\cite{Ost02,angel05} was placed 
at a distance of 37 mm and 42 mm from the
0.950 mg/cm$^2$ and 2.080 mg/cm$^2$ thick $^{208}$Pb targets respectively.
The DINEX telescope
covered therefore different laboratory angles  for the different targets ranging
from 136.3$^{\circ }$ to 166.6$^{\circ }$ for the thin target and from 
131.8$^{\circ }$ to 164.5$^{\circ }$ for the thick target of 2.080 mg/cm$^2$ 
used only for the 22 MeV $^6$He beam. 
The DINEX array consisted of four quadrants forming a CD \cite{Ost02}, 
each composed of single sided Si strip detectors
($\Delta $E) 40 $\mu $m thick with sixteen radial strips stacked in
a 500 $\mu $m thick single PAD Si-detector. Each strip subtended an angle 
of about 2$^{\circ}$, although this value depends on the scattering angle.


The energy calibration was performed using a triple alpha source 
for the front single sided Si strip detector of the telescope. 
The telescope as a whole was calibrated using the elastic scattering peaks of the $^4$He and $^6$He 
beams from the $^{208}$Pb target at 
different energies below the Coulomb barrier
where the elastic peak at backward angles still has significant statistics.
To evaluate the energy losses of the beam and ejectiles
in the different media, i.e. target thickness and dead layers of the front and back detectors,
a simulation programme developed by R. Raabe~\cite{rab01} 
and adapted to the geometry of our setup was used. 
The energy losses of the $^{4,6}$He ions were calculated in tables using SRIM~\cite{SRIM}
and inserted into the simulation programme.
The fact that the interaction region had a finite size, was also taken into account in the simulations.

In order  to obtain the energy of the ejectile we first added the signals 
from the two detectors of the telescope, $\Delta $E and E. The latter
was multiplied by a matching constant $\alpha(\theta ) $ 
that depends on the relative gain of the different
electronic chains (see~\cite{angel06} for more details). 
Then we used the energy of the elastic peaks of $^{4}$He at 12 MeV 
and $^{6}$He at 14, 16 and 18 MeV to calibrate the total  
telescope signal.

\begin{figure}
 {\centering \resizebox*{0.7\columnwidth}{!}
 {\includegraphics{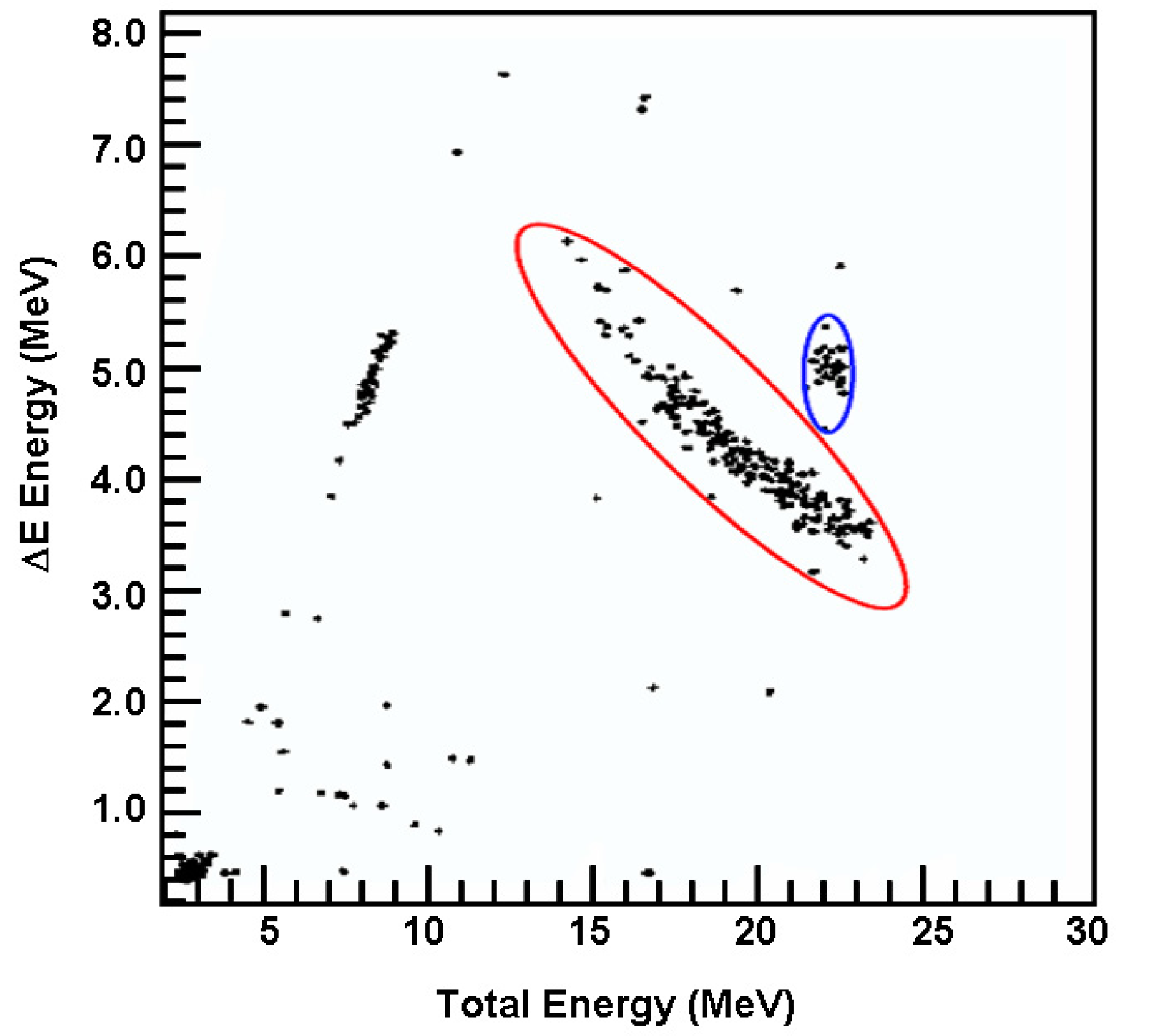}} \par}
 \caption{\label{bidim}
Two dimensional plot of the  22 MeV $^6$He scattered from a 2.08 mg/cm$^2$ $^{208}$Pb target 
at laboratory angle of 144(1)$^{\circ }$. The regions enclosed by solid lines correspond to the $^6$He and 
$^4$He events. The plot  includes all events in coincidences with multiplicity $\geq 1$ to 
stress the low background level. Therefore one can see events fully stopped 
in the $\Delta E$ detector corresponding to $\alpha$'s from the grand-daughter of $^{210}$Pb 
and $^{234}$Th present in the target in very small amount. The count rate was 
less than two $\alpha$ particles every 10 min for this angular coverage. }
 \end{figure}

The DINEX telescopes allow mass and charge separation
of the reaction products. A typical mass spectrum, 
$\Delta $E versus total energy, $E_T$, 
obtained with the DINEX array for a $^6$He beam energy of 22.0(1) MeV 
and $\theta_\mathrm{lab} $ = $144^{\circ }$ $\pm $ 1$^{\circ }$ is shown 
in Fig.~\ref{bidim}.  Using these calibrations and the mass and charge separation obtained with the telescopes
we are able to identify the energy distribution
of the breakup products as well as the elastic counts in every ring ($\theta_i$), as shown 
in this figure. The energy loss in the 40 $\mu$m thick $\Delta$E detector is displayed 
versus total energy. The elipsoids select the elastic and
breakup events  used in the analysis. The low energy selection corresponds to protons. 
The background events are remarkably low considering that no condition beyond coincidence between the front and back detectors
of the telescope is applied.

\begin{figure}
 {\centering \resizebox*{0.75\columnwidth}{!}
 {\includegraphics{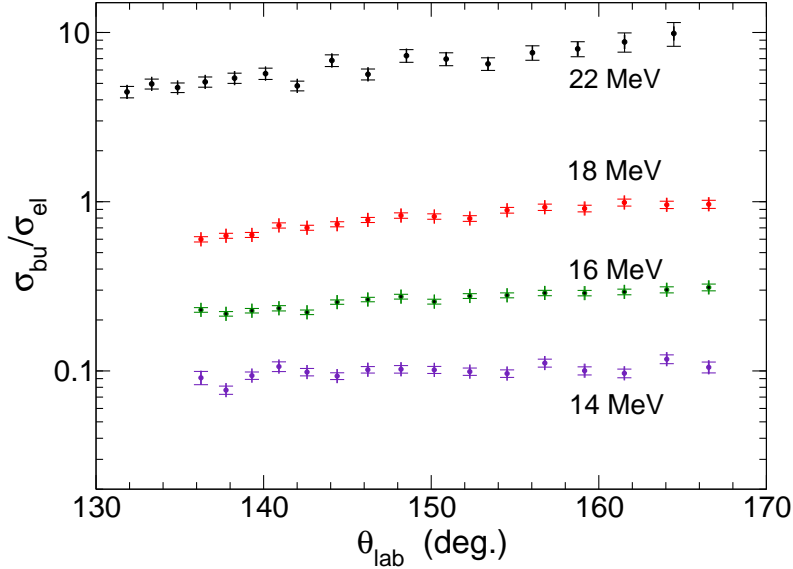}} \par}
 \caption{\label{PEang}
  Ratio between the measured $^4$He and $^6$He events, as a function of the 
  laboratory scattering angle, at several bombarding energies. The angular range
at 22 MeV is different to that covered at energies below the
Coulomb barrier due to different positioning of the target. See section 2 for more details. Errors 
bars correspond to statistical errors.}
  \end{figure}

We obtained the breakup cross section by making use of the ratio of breakup to elastic events seen in
the telescopes. To have reasonable statistics, we grouped the breakup events in 1 MeV bins. Then, we evaluated for each
detector strip $\theta_i$ and for each energy bin $E_i$ the ratio of breakup to elastic events $N_{bu}(\theta_i, E_i)/N_{el}(\theta_i)$.
Note that uncertainties associated with beam intensity, solid angle, efficiency of the electronic chain or target thickness disappear
in this ratio. When we add these ratios for all the energy bins, we obtain the ratio of breakup to elastic cross sections as a function of the scattering angle. These ratios are shown in Fig.~\ref{PEang},
as a function of the laboratory scattering (LAB) angle , at several  beam energies. 
It is noticeable that, at $E_\mathrm{lab}=22$~MeV, the yield of $\alpha$ fragments exceeds 
the elastic ones  by a 
factor of ten. As discussed in the introduction, this
large $\alpha$ cross section  has been reported  for other reactions induced by 
\nuc{6}{He} on several medium-heavy targets, such as  \nuc{65}{Cu} 
\cite{Nav04},  \nuc{64}{Zn} \cite{DiPiet04}, and \nuc{209}{Bi}  \cite{Agu00}.

The breakup double differential cross section, with respect to the angle and the energy of the $\alpha$ particle,
depends on energy and angle, and can be related to the ratio of the number of counts and to the elastic differential cross
section by the following expression:

\begin{equation}
\frac{d^2\sigma_{bu}}{dEd\Omega}\Big{|}_{E_j,\theta_i} = 
\frac{N_{bu}(\theta_i,E_j)}{N_{el}(\theta _i)}\frac{1}{\Delta E_j}\left( \frac{d\sigma_{el}}{d\Omega}\right)_{\theta _i} 
\label{eq:sec1} 
\end{equation}
where $\Delta E_j=1$~MeV is the bin width. The differential elastic cross section for the different 
laboratory angles are taken from~\cite{angel06,tesisangel}.

\begin{figure}
 {\centering \resizebox*{0.7\columnwidth}{!}
 {\includegraphics{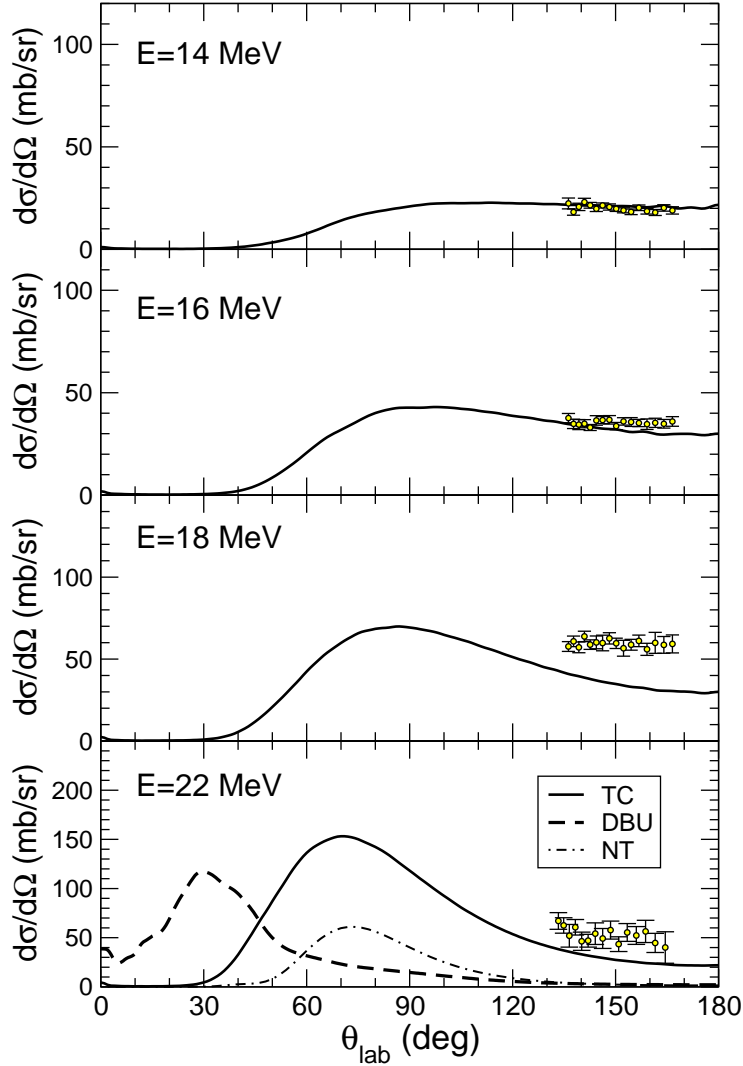}} \par}
 \caption{\label{he4ang}
  Angular distribution of $\alpha$ particles arising from $^{6}$He fragmentation, 
  in the laboratory frame, for several incident energies. Experimental angular distributions are
compared with transfer to the continuum (TC) calculations (solid lines). The distribution obtained at
 22 MeV is also compared with a direct breakup (DBU) calculation, performed within the CDCC approach 
(dashed line), and a DWBA calculation for 
the one neutron transfer (NT) 
leading to bound states of the $^{209}$Pb nucleus (dotted-dashed line).}
  \end{figure}

If this expression is integrated over the energy of the breakup fragment, one obtains the  differential 
breakup cross section,
as a function of the scattering angle, which is given by:
\begin{equation}
\frac{d\sigma_{bu}}{d\Omega}\Big{|}_{\theta_i} \approx \sum_{j=E_{min}}^{j=E_{max}} 
\frac{N_{bu}(\theta_i,E_j)}{N_{el}(\theta _i)}\left( \frac{d\sigma_{el}}{d\Omega}\right)_{\theta _i} .
\label{eq:sec4b} 
\end{equation}
These differential cross sections are presented in Fig.~\ref{he4ang}. Errors bars correspond to statistical errors. It should be noted that the breakup differential cross sections
are highest around the Coulomb barrier ($E= 18$ MeV), and they decrease both at lower energies and higher energies. However, for
14 MeV, which is well below the barrier, the breakup cross sections are still sizeable. This indicates that the mechanism
producing alpha particles is effective for projectile-target separations as large as 17.3 fm, which corresponds to 
the distance of closest 
approach for a head-on collision at this energy. This mechanism, indeed, will be
a source of the long range absorption which we have seen in the analysis of elastic data.

\begin{figure}
 {\centering \resizebox*{0.75\columnwidth}{!}
 {\includegraphics{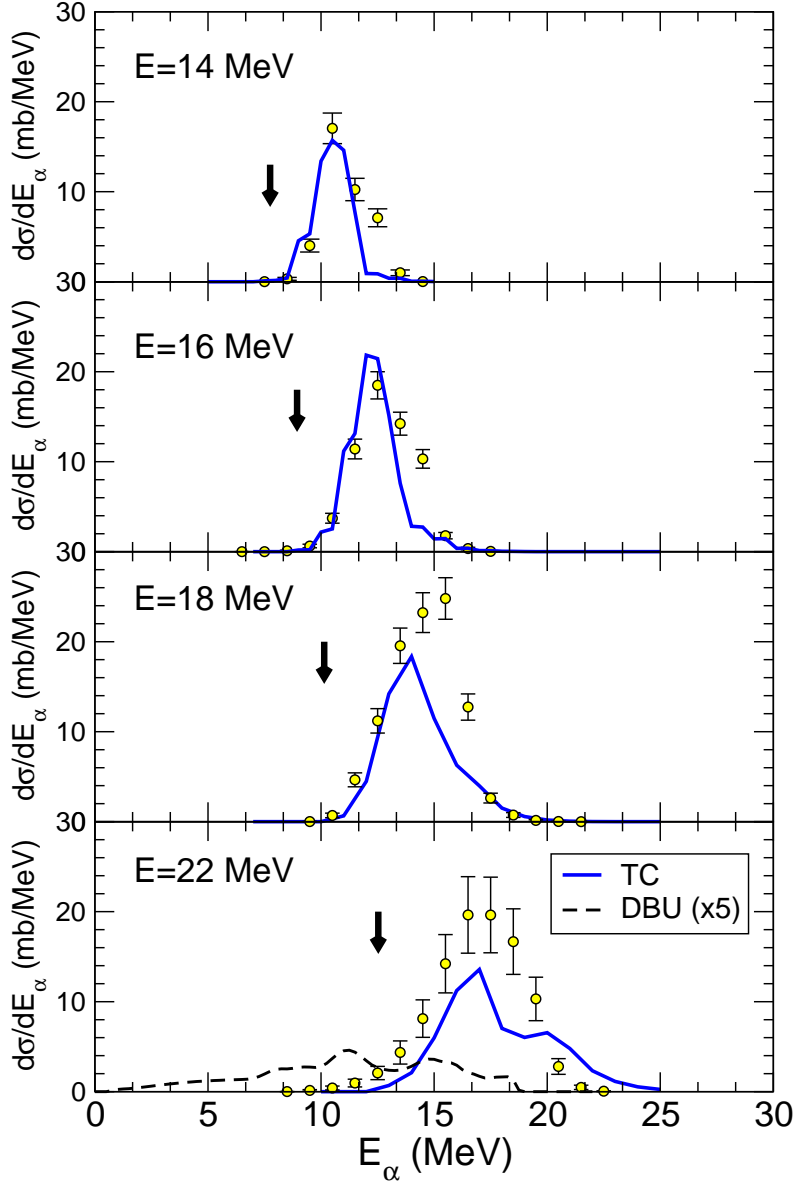}} \par}
 \caption{\label{he4vsE}
  Energy distribution of detected $\alpha$ particles for the reaction
  \nuc{6}{He}+\nuc{208}{Pb} at 14, 16, 18 and 22 MeV, integrated in 
  the angular range 132$^{\circ}$-164$^{\circ}$. 
  Experimental data (circles) are compared with transfer to the continuum (TC) calculations (solid lines). The arrows
  indicate the expected energy of the $\alpha$ particles scattered at 
  $\theta_\mathrm{lab}=151^\circ$, assuming an extreme direct breakup picture. 
  In the panel for $E_\mathrm{lab}=22$ MeV the full direct breakup calculation, performed within the 
  CDCC method,  is also presented, 
 multiplied by a factor of 5. 
   }
  \end{figure}

Similarly, if, for each energy bin of the $\alpha$ particles,
the double differential cross sections are integrated with respect to the angle, over the angular range covered by the CD detectors,
one gets a breakup differential cross section as a function of the energy, which is given by
\begin{equation}
\frac{d\sigma_{bu}}{dE}\Big{|}_{E_j} \approx \sum_{i=1}^{i=16} 
2\pi \sin(\theta_i )\frac{\Delta \theta_i}{\Delta E_j}
\frac{N_{bu}(\theta_i,E_j)}{N_{el}(\theta_i)}
\left( \frac{d\sigma_{el}}{d\Omega}\right)_{\theta _i}  .
\label{eq:sec3} 
\end{equation}
These cross sections are presented in Fig.~\ref{he4vsE}.

Our purpose is to understand which  mechanism is responsible for the large production of $\alpha$ particles.
The fact that the $\alpha$ particles are produced with relatively large energies, that increase with the 
projectile energy, leads us to conclude that the process should be a direct one, and not a compound nucleus 
formation. Within the direct mechanism picture, we can consider three mechanisms:

a) Transfer to the continuum: As the $^6$He nucleus gets close to the target, it leaves the two neutrons with low kinetic energy 
 with respect to the $^{208}$Pb target, and the remaining $\alpha$ particle escapes. If this is the case, we
 would expect that 
the $\alpha$ particle would have an energy distribution centered around the energy of the elastically scattered $^6$He.

b) Direct breakup: The  $^6$He nucleus breaks up in the field of the target and it goes to a continuum state with low excitation energy.
In this case, we would expect that the  $\alpha$ particle (and the neutrons) to have a similar velocity to the elastically 
scattered  $^6$He, and hence its energy would have a broad distribution around 4/6 of the energy of $^6$He.

c) Neutron transfer: One of the neutrons of  $^6$He is transferred to the target, producing a bound state of  $^{209}$Pb and leaving
$^5$He in a broad resonance, that  rapidly decays producing $^4$He. In this case, the kinetic energy of the $^5$He resonance, although
dependent on the Q-value, would be similar to that of the elastically scattered   $^6$He (for $Q\simeq 0$), and the alpha particles would have
a broad distribution around  4/5 of the energy of $^5$He.

It should be noted that, in our work,  direct breakup, 2n transfer and 1n transfer correspond to 
different approaches to describe the mechanism that produce alpha particles, rather than to 
different reaction channels.  From the theoretical point of view, these three  mechanisms describe the 
removal of the valence neutrons in \nuc{6}{He},  but they do not lead necessarily to different final 
states.  They 
should be seen as different approaches to a very difficult 4-body problem ($\alpha$ + n+ n+ \nuc{208}{Pb}) 
which 
cannot be solved  accurately. 
Each of
these methods  emphasize a different way in which the fragmentation is produced. In the 
direct breakup method, one assumes that the \nuc{6}{He} is broken up by exciting the neutrons to 
continuum states with low relative energy with respect to the $\alpha$ core,  
and hence a representation in terms of the \nuc{6}{He} continuum states is used. In the transfer to the 
continuum approach, it is assumed that the fragmentation occurs by transfer of the  valence neutrons to 
weakly bound states of the target, and hence a target representation is preferred in this case. Finally, 
the 1n transfer corresponds to an intermediate situation, in which one assumes that one of the neutrons 
is transferred to states of low  relative energy with respect to the target, while the other 
remains in a low energy state  with respect to the alpha core. 
Then, these approaches should be understood as extreme pictures that emphasize different degrees of freedom of the breakup process, and 
not different reaction channels. For instance, 
it has been shown \cite{Mor06a} that, if a large basis is included in both 
the  direct breakup and transfer to the continuum
representations, there is a strong overlap between the states populated in both methods. As a consequence, 
the cross sections calculated within the two approaches can not be simply added  to obtain the total 
breakup cross section.

It is apparent from Fig.~\ref{he4vsE} that
the position of the energy peak is not consistent with the expected value in a direct breakup
picture, in which the fragments are produced with essentially the beam velocity. The results for this estimate, calculated 
at $\theta_\mathrm{lab}=151^\circ$, are shown by  the arrows in Fig.~\ref{he4vsE}.  Notice 
that these values are significantly 
smaller than the measured energy of the $\alpha$ particles and hence it is not 
expected that the direct breakup model is suitable for understanding the present data.

Under the assumption that
the relative energy between the halo neutrons remains small during the process,
the gain of kinetic energy of the outgoing $\alpha$ particles   
implies that these  neutrons are left with a small 
(or even negative, if they are 
transferred to bound states)
relative energy with respect to the target. Energy 
conservation demands  that the available kinetic energy is used to 
excite the target, or to accelerate the  $\alpha$  particles, as we observe 
in this experiment. This indicates that the processes responsible 
for the production of these   $\alpha$  fragments 
are of  a more complicated nature than suggested by the simple direct breakup model.

From these semiquantitative considerations, 
our data suggest a transfer to the continuum picture, in which the valence neutrons are 
transferred to highly excited states of the target, lying 
around  the \nuc{210}{Pb}$\rightarrow$ \nuc{208}{Pb}+2n breakup threshold. Final 
states above this threshold can be interpreted as a three-body breakup of the projectile,
while
those below the threshold would correspond to a pure transfer process leading  to bound states of 
the \nuc{210}{Pb} residual nucleus.

\section{Theoretical calculations}

In this section, we present a more quantitative analysis of the data, by performing calculations 
for the different mechanisms described in the preceeding section (transfer to the continuum, direct breakup or one neutron transfer). The goal of this study is to see which of these mechanisms is more appropriate 
to describe the present data. 


\subsection{\label{sec:tc} Transfer to the continuum }

\begin{figure}[h]
 \vspace{0.3cm}
  \hfill
  \begin{minipage}[t]{.45\textwidth}
       \par (a)
    \begin{center}  
      \epsfig{file=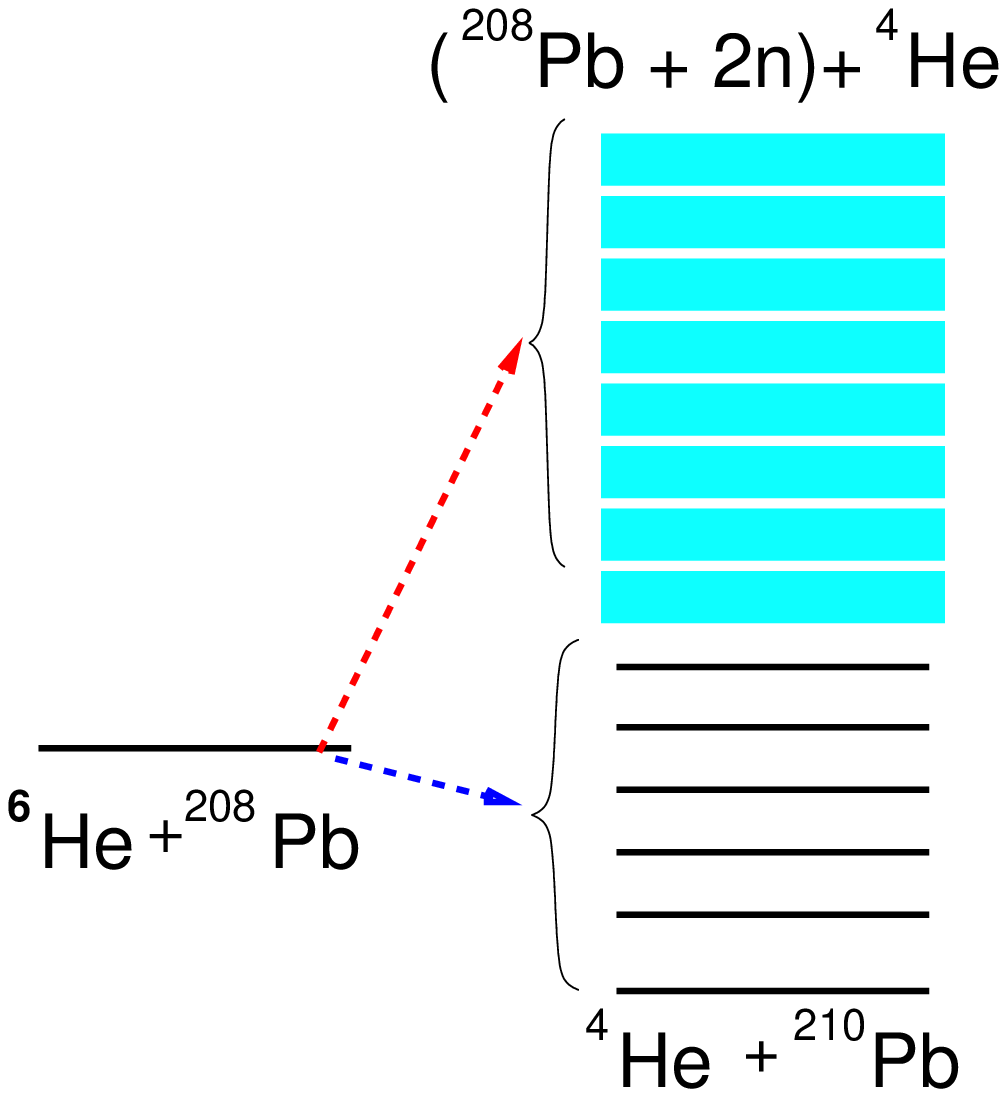, scale=0.5}
    \end{center}
  \end{minipage}
  \hfill
  \begin{minipage}[t]{.5\textwidth}
     \par (b)   
    \begin{center}
      \epsfig{file=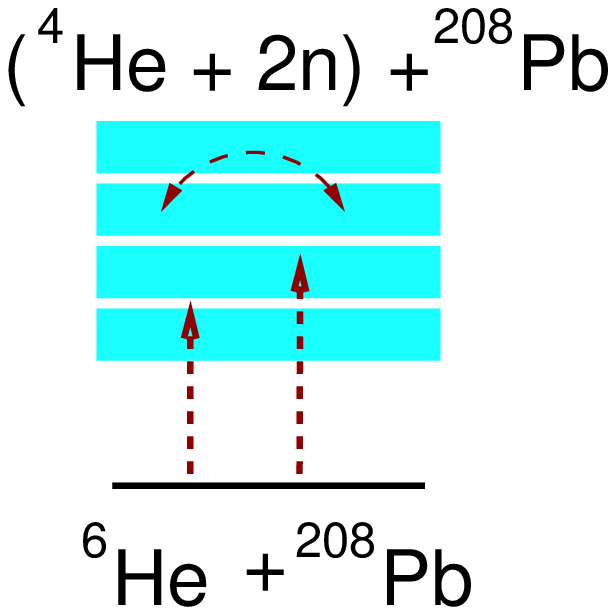, scale=0.5}
    \end{center}
  \end{minipage}
  \caption{\label{Fig:couplings}  Schematic representation of the couplings included in the 
            transfer to the continuum (a) and direct breakup  (b) calculations.}
  \hfill
\end{figure}

We use the  transfer to the continuum method, in which the  
two-neutron removal is treated as a transfer  of the valence neutrons to bound 
and unbound states of the 2n+\nuc{208}{Pb} system. For simplicity, these calculations 
are performed within the DWBA approximation. 
Although  the DWBA method has been traditionally applied to the transfer between bound states, it has 
also proved to be a useful method in situations where final states lie in the continuum 
\cite{Vin70,Fro79,Baur76,Uda89,Mor03a}.
To simplify our description of the reaction process, we assume that the 
relative motion of the two halo neutrons is not affected during the process. At 
least for the Coulomb interaction, which is 
known to be very important in this reaction, this assumption is expected to 
 be reasonable because
this interaction will act only on the $\alpha$ fragment, 
tending to stretch the \nuc{6}{He} system,
with the neutrons moving against the $\alpha$ core. So, in these calculations, we 
emphasize the dineutron-$\alpha$ relative coordinate, under the assumption that this is 
the \nuc{6}{He} main degree of freedom which is excited during the process. 
These 
transfer couplings are schematically depicted in 
Fig.~\ref{Fig:couplings}a.

In this reaction, the post form expression of the DWBA transition amplitude involves a 
matrix element of the operator 
$$V_\mathrm{[{2n}-\alpha]}+U_\mathrm{[\alpha-\nuc{208}{Pb}]}-
U_\mathrm{[\alpha-\nuc{210}{Pb}]}. $$  
The potential parameters required in the calculations are 
summarized in Table \ref{omp}. For the $\alpha$-\nuc{208}{Pb} system, we used 
the parameterization of  Barnett and Lilley \cite{Bar74}.  The optical potential for the 
\nuc{6}{He}-\nuc{208}{Pb}
system, which is used to generate the distorted waves for the incoming channel, is taken 
from a fit of the elastic angular distribution \cite{angel06,tesisangel}.  The same parameters were 
used for the outgoing channel ($\alpha$-\nuc{210}{Pb}). 
As explained above, the neutron pair was allowed to be 
transferred to both  bound and unbound states of \nuc{210}{Pb}. The
2n-\nuc{208}{Pb} relative wavefunctions were generated with the 
deuteron-\nuc{208}{Pb} optical potential derived in Ref.~\cite{DCV80}. 
In order to permit the inclusion of bound states, only the 
real part of this potential was considered. 
Reduced radii ($r_x$) were converted to 
physical radii ($R_x$) as $R_x=r_x (A_1^{1/3} + A_2^{1/3})$, for the \nuc{6}{He}-\nuc{208}{Pb}
and $\alpha$-\nuc{210}{Pb} systems, and as  $R_x=r_x$ for the \nuc{}{2n}-$\alpha$ system.

\begin{table}
\caption{\label{omp}Potential parameters used in the  calculations.}
\vspace{0.3cm} 
\begin{tabular}{cccccccc}
\hline
System                 &    $V_0$  &   $r_0$   &  $a_0$ &  $W_0$  & $r_i$  & $a_i$  & Ref. \\   
                       &    (MeV)  &    (fm)   &  (fm) &   (MeV) &  (fm)  &  (fm)  &       \\
\hline
\nuc{6}{He}-\nuc{208}{Pb}$^a$ &  
                               $b$ &   1.015  &  1.15  &   $c$  &   1.015   &  1.70  &  \cite{angel06} \\ 
$\alpha$-\nuc{208}{Pb} &   96.44    & 1.376 & 0.625 & 32 & 1.216 & 0.42  & \cite{Bar74} \\
{2n}-\nuc{208}{Pb}  &  85.55 & 1.20  & 0.751 & -  &   -   &   -   &  \cite{DCV80}  \\
{2n}-$\alpha$	  &  87.18 & 1.90  & 0.39  & -  &   -   &   -   &   \cite{Rus01} \\
\hline
\end{tabular}
\vspace{0.3cm}

$^a$ The same parameters were used for the \nuc{4}{He}-\nuc{210}{Pb} distorted potential. \\
$^b$ {$V_0$= 32.8 31.4, 33.1 and 5.89 MeV for $E_\mathrm{lab}=$14, 16, 18 and 22 MeV, respectively.}\\
$^c$ {$W_0$=0.04, 4.6, 5.1 and  9.84 for $E_\mathrm{lab}=$14, 16, 18 and 22 MeV, respectively.}

\end{table}


In the di-neutron model, the \nuc{208}{Pb}+2n final states should be considered as {\em doorway states} to 
which the di-neutron is transferred. The {\em doorway states} subsequently fragment into bound or continuum states of \nuc{210}{Pb}. To 
evaluate the wavefunctions of these {\em doorway states}, we take into account the average separation energy of the 2n single particle configurations for each $L$ value, 
$\langle \epsilon_\mathrm{2n} \rangle$, according to the available experimental 
information \cite{pb210}. 
Then, we evaluate the number of nodes of the \nuc{208}{Pb}-2n relative 
wavefunction, $N$,  
preserving the Pauli principle. This is done using the  Wildermuth condition \cite{Wil66,Sat83}: 
$2(N-1)+L = 2(n_1-1)+l_1 + 2(n_2-1)+l_2$, where $(n_1,l_1)$ and $(n_2,l_2)$ are the single-particle 
configurations which have to be populated in a simple shell model picture to produce a state with 
the desired $J$.  In all cases, 
we used a Woods-Saxon form, with radius 
1.2$\times$208$^{1/3}$ fm and diffuseness $a=0.75$~fm \cite{DCV80}.
Finally, we adjust the potential depth, $V_0$, to produce a state with the given $L$, $N$,
 and  $\langle \epsilon_\mathrm{2n} \rangle$. The case $L=1$
deserves a special consideration, since no single-particle pair configuration of bound single-particle states couples to 
$J=1^-$. To obtain these states one neutron has to be promoted to the next shell, and this will increase the 
energy of the $1^-$ state by $\hbar \omega \simeq  $ 6 MeV, leading to a resonance around 
$\langle \epsilon_\mathrm{2n} \rangle$=1~MeV, above the two-neutron breakup threshold.  Then, the potential 
depth was adjusted in this case to obtain a resonance at this energy. The values of $N$, $V_0$ and $\langle \epsilon_\mathrm{2n} \rangle$ are listed in Table  \ref{binding}.

\begin{table}
\caption{\label{binding} 2n-\nuc{208}{Pb} potentials used to generate the bound and unbound states of the $^{210}$Pb nucleus
in the transfer to the continuum calculations.}
\vspace{0.3cm}
\begin{center}
\begin{tabular}{lccccccc}
\hline
$J^\pi$                 &    $0^+$  &   $1^-$   &  $2^+$  &  $3^-$  & $4^+$   & $5^-$  & $6^+$ \\
\hline
$V_0$ (MeV)            &   85.9    &   73.2      &   85.6  & 100.3   &  85.6   & 97.3   & 85.1   \\
$N         $              &      7    &     7     &   6     &    6    &    5    &  5     &  4     \\
$\langle 
\epsilon_{2n}
\rangle$ (MeV)         & -4.9   &   1.0      &  -4.8      &  -6.4    & -5.0   & -5.1       &  -5.4      \\
\hline
\end{tabular}
\end{center}
\end{table}

The states of  \nuc{210}{Pb} above the \nuc{208}{Pb}+2n threshold are described by means of 1~MeV continuum bins, 
obtained by superposition of the 2n-\nuc{208}{Pb} scattering wavefunctions, up to a 
maximum excitation energy of 8 MeV. Thus, the two-body spectroscopic strength is naturally distributed 
along the continuum. 
 However, for bound states, the fragmentation does not appear. To take this into account, we consider that the {\em doorway states} are fragmented into $N_b$ states, with two neutron 
separation energies of $S_{2n}$=0.5, 1.5,$\ldots$,7.5 MeV, each one with an spectroscopic factor of $1/N_b$. In 
practice, we took $N_b = 8$ but we verified that the results did not depend strongly on this number. For  the continuum 
states, we assumed unit spectroscopic factors.

The $\alpha$-2n interaction, required to generate the \nuc{6}{He} ground state 
wavefunction, was parameterized using a standard Woods-Saxon 
form, with radius $R_0=1.90$~fm and diffuseness $a_i=0.39$~fm, which corresponds to the 
set II of Ref.~\cite{Rus01}. 
A pure $2S$ configuration, with unit spectroscopic factor, was
assumed for this state. The potential depth was obtained using the energy 
separation method, 
that is,  the depth  was adjusted in order to 
reproduce the two-neutron separation energy.  However, instead
of using  the experimental separation energy  ($S_{2n}=0.975$ MeV) we 
used the modified
value $S_{2n}=1.5$~MeV. This change is motivated by the fact that the 2n-$\alpha$ wavefunction, 
calculated  with the experimental separation energy, extends too much in configuration space, as 
compared to a realistic three-body calculation. As a consequence, couplings to the continuum 
are largely overestimated, leading to unrealistic results for the scattering observables, as
shown recently for the elastic scattering of 
$^6$He+$^{208}$Pb  \cite{Rus05} and  $^6$He+$^{209}$Bi \cite{Mat06a}.  By 
increasing the binding energy to $S_{2n}=1.5$~MeV the wavefunction obtained in the di-neutron model 
simulates fairly well the three-body wavefunction  in the (nn)-$\alpha$ coordinate. The physical idea 
behind this choice  is that, in \nuc{6}{He}, the  
neutron-neutron pair contributes to the binding energy with a positive average 
energy which, added to the (negative) relative energy 
associated to the  (nn)-$\alpha$ motion, should give 
the correct binding energy.
Further details of this method will be published elsewhere \cite{din06}.
%

In order to get convergence of the angular cross section within the angular 
range covered by the present data, we found it necessary to include partial waves
up to $L=6$ for the 2n-\nuc{208}{Pb} motion. The total angular momentum was 
set to $J=50$, and the distorted waves were calculated up to 200 fm. A range of 
non-locality of 9 fm was required for the transfer couplings. These 
calculations were  performed with the coupled-channels 
computer code FRESCO \cite{fresco}.

The differential angular distribution for each final state is proportional to 
the square of the DWBA amplitude. By energy conservation, the energy 
of the outgoing $\alpha$ particles is obtained from  the excitation energy 
of  \nuc{210}{Pb}. This procedure provides 
a double differential cross section with respect to the angle 
and energy of the scattered $\alpha$ fragments. In order to permit a 
meaningful comparison with the data, these 
magnitudes were transformed to the LAB frame. For this purpose, the
calculated center-of-mass (CM) double-differential cross sections were converted to LAB system
using the appropriate Jacobian for the kinematical transformation for 
$(\theta_\mathrm{c.m.}, E_{\alpha}^\mathrm{c.m.})$ $\rightarrow$ 
$(\theta_\mathrm{lab}, E_{\alpha}^\mathrm{lab})$.

The calculated angular distributions are represented by the solid lines 
in Fig.~\ref{he4ang}. The overall agreement with the data is good, although for
the higher scattering energies ($E_\mathrm{lab}=$18 and 22 MeV) the experimental distributions
are somewhat underpredicted. This underestimation, which could be due to the approximations
involved in our method, might indicate the presence of other 
channels not included in our calculations. Dynamical effects, such as multi-step transfer processes,
not considered in the DWBA calculations, could also affect the results.  We would like to note that 
the calculation of absolute cross sections in two-neutron transfer reactions is a very complicated 
problem \cite{Fen76b,Bay82},  and discrepancies as large as  one or two orders of magnitude 
between theory and 
experiment have been reported by some authors. Keeping in mind these difficulties, the degree of 
agreement between the present data and the transfer to the continuum calculations is
very encouraging. 

Finally, the calculated energy distributions of the $\alpha$ particles, in the LAB frame,
are given by the solid lines in Fig.~\ref{he4vsE}. The transfer to the continuum calculations (thick solid lines)
 reproduce 
very well the shape of these distributions. In particular, 
the position of the peak is very well accounted for at all energies. The absolute values 
of these distributions are also reasonably reproduced, except for the underestimation 
at the higher energies discussed above.

 We would like to  stress that these calculations do not include any free parameter. They are
 based on a direct application of a fully quantum mechanical expression of the transition amplitude, within 
 the DWBA approximation, and the ingredients are the potentials between the fragments taken 
 from the literature, and a physically motivated model for the initial and final states in the
 \nuc{6}{He} and \nuc{210}{Pb} nuclei, respectively.

\subsection{Direct breakup \label{sec:cdcc} }
The direct breakup component of the $\alpha$ inclusive spectrum could be also calculated within 
the standard continuum discretized coupled-channels (CDCC) method \cite{Aus87}. In the direct breakup picture, the fragmentation process is formally treated as an inelastic 
excitation of the projectile to the continuum (see Fig.~\ref{Fig:couplings}b).  From the 
semiquantitative arguments outlined in Sec.~\ref{sec:general}, we do not expect this scheme to 
be appropriate for the present reaction, since the observed energy of the $\alpha$ particles is
significantly larger than the values estimated by kinematic considerations assuming a 
direct breakup picture. These considerations, along with  the calculations presented in Sec.~\ref{sec:tc}, 
clearly suggest that  
the  energies of the observed $\alpha$ particles at backward angles are consistent 
with the transfer of the valence neutrons to weakly bound states of the target. These states are indeed better described in 
a basis of the target representation, as we have done in the transfer to the continuum 
calculations. The direct breakup representation, by contrast, is expected to be less efficient in this case, in the 
sense that a large basis would be required to describe 2n-target states with small relative energy and 
angular momentum. In order to test these arguments we have performed CDCC calculations for the 
$E_\mathrm{lab}=22$ MeV case.


Again, 
we assumed a simple di-neutron model for \nuc{6}{He}. 
Partial waves $s$, $p$ and $d$  were included for the 2n-$^4$He 
relative motion. For each partial wave, 
the  $^6$He continuum was divided into energy bins,  
according to the scheme detailed in \cite{Rus05}.  In analogy with the prescription used in
the transfer to the continuum calculations, the effective two-neutron separation energy $S_{2n}=1.5$~MeV was 
used to generate the $^6$He ground state and continuum wavefunctions. With this prescription, 
the elastic angular distribution is very well reproduced. 
The breakup angular and energy distributions obtained from this 
calculation are represented in Figs.~\ref{he4ang} and \ref{he4vsE} by the dashed 
line (bottom panel). It can be seen that 
the experimental data are underestimated by almost an order of magnitude, at all the
measured angles. The inclusion of higher partial waves did not solve the discrepancy.  Hence, as we 
anticipated, the direct breakup picture is inadequate to describe the present data. From 
the figure, we see also that the CDCC distribution dominates the small angle region, 
with a pronounced peak  
around 30$^\circ$. At these angles, the transfer to the continuum curve is very small, suggesting that for the description of the 
$\alpha$ particles emitted at forward angles the direct breakup scheme should provide a more suitable
representation than the TC.

We note that a proper CDCC calculation for the present reaction would require a three-body description 
 of \nuc{6}{He}, thus giving rise to a four-body scattering problem.  This  kind of calculations  
has been recently reported by the Kyushu group 
for $^{12}$C \cite{Mat04} and $^{209}$Bi \cite{Mat06a} targets. Although 
our
results will be modified to some extent if a realistic  
three-body description  of \nuc{6}{He} is used, we believe that our simplified 
calculations retain the essential physics to illustrate the inadequacy of the CDCC method to 
describe the present data.




\subsection{One neutron transfer  \label{sec:1n} }
Another channel that could contribute to the production of $\alpha$ particles is the 
one neutron stripping, $^{208}$Pb($^{6}$He,$^5$He)$^{209}$Pb. This process will 
produce   $^5$He in a resonant state, which will eventually 
decay into n+$^4$He.  The calculations for this process 
were performed within the DWBA approximation.  Due to Q-value considerations, this process 
populates mainly 
bound states of the $^{209}$Pb nucleus, similar to what occurs in the 
$^{208}$Pb(d,p)$^{209}$Pb reaction \cite{Kov74}.
 Thus, we included the known 
bound states for the $^{209}$Pb nucleus, and assumed unit spectroscopic factors, since 
these  are known to be mainly single-particle states. Concerning  the spectroscopic 
factor for  $^6$He $\rightarrow$ $^5$He+n, in a strict
shell model picture, with maximal pairing,
spectroscopic factor for neutron transfer is equal  to number of valence neutrons, what would 
give 2. We adopt however the value 1.60 for this spectroscopic factor, reported by 
Nemets {\em et al} \cite{Nem88}, 
and obtained using the method detailed in Ref.~\cite{Smi77}. 


For the entrance channel,  $^6$He+$^{208}$Pb, the effective 
potential obtained from CDCC calculations 
was used. For  the exit channel,  $^{5}$He+$^{209}$Pb, the potential
was calculated by folding the n+$^{209}$Pb potential from the compilation 
of Varner {\em et al.} \cite{Var91} and the $\alpha$ + $^{208}$Pb potential
of Goldring {\em et al.} \cite{Gol70}
with the ground state wave function of $^5$He. The latter was  represented by an 
energy bin of a width 1.2 MeV placed at an excitation energy of 0.8 MeV
above the $^5$He$\rightarrow ^4$He+n breakup threshold.
The binding potential for
$^4$He+n was taken from Ref.~\cite{Ban76}.

The calculated angular distribution is shown by the dotted-dashed  line in the lowest panel
 of Fig.~\ref{he4ang}  ($E_\mathrm{lab}$=22 MeV). 
It should be  noted that, in this case, the scattering angle corresponds to the center of mass of the 
\nuc{5}{He}* system, rather than the \nuc{4}{He} angle.  At intermediate angles ($\theta_\mathrm{lab} 
\approx 90^\circ$), our 
calculation predicts a significant contribution of the 1n transfer, although the $\alpha$ yield is still 
dominated by the 2n transfer. This result is 
in agreement with the experimental data of De Young {\em et al.} \cite{DeY05} in which the 2n transfer 
was identified as the main mechanism producing $\alpha$ particles at these angles, with a small contribution 
arising from 1n transfer. 
We note however that the  cross sections   
reported in \cite{DeY05} for the 1n and 2n transfer processes 
are inferred from the $\alpha$ particles observed at 90 and 120 degrees,  while in our experiment 
the observed $\alpha$ particles  correspond to larger angles (above 130$^\circ$). 
Therefore, the cross sections reported in  Ref.~\cite{DeY05}  cannot be readily extrapolated to 
our case.  Moreover, even if we compare the same angular range, we do not 
expect to get the  same quantitative results, because the  
states populated in both reactions (namely, $^{209}$Pb and $^{210}$Bi)  are  different. 

It can also be seen that at the angles of interest of our 
experiment ($\theta_\mathrm{lab} > 130^\circ$), the contribution of the 1n transfer is negligible 
and  the production of $\alpha$ particles  is essentially due to 2n transfer. Moreover, the 1n 
transfer cannot explain the  energy of the $\alpha$ particles since    
the maximum kinetic energy of the $\alpha$ particles emitted in this process, which corresponds to 
a transfer to the  $^{209}$Pb ground state, is about 17.5 MeV, while the experimental energy 
distribution extends beyond 20 MeV. 
Therefore, we conclude that the one neutron stripping channel is not responsible for the 
underestimation of the data at these angles.



\section{\label{sec:discussion} Discussion}

The results found in this work are consistent and complementary to our 
previous analysis of the elastic scattering for the same reaction \cite{angel06}. The 
optical model analysis performed in the previous work revealed the importance of a
long range absorption mechanism. With the present analysis, one can conclude 
that this mechanism is presumably related to the two-neutron removal process discussed in this work. In 
order to draw more  definite conclusions it would be 
desirable to obtain data  in a wider angular range. 

Furthermore, in order to understand more clearly the importance of 
different reaction mechanisms, it would be very useful  to perform a similar 
experiment in which the neutrons could be detected in coincidence with the 
$\alpha$ particles. Given the small detection efficiency for neutrons, 
and the fact that the two neutrons can be emitted, in principle, in arbitrary 
directions, this experiment is also very challenging from the experimental 
as well as from the theoretical points of view. Despite these difficulties, a similar
experiment has been recently performed for the reaction 
\nuc{6}{He}+\nuc{209}{Bi} at 22 MeV \cite{DeY05}. By measuring neutron-$\alpha$ 
 coincidences it is  concluded that  approximately 55\% of the observed $\alpha$ yield 
around and beyond the grazing angle is due to two-neutron transfer to unbound states of 
the \nuc{211}{Bi} nucleus.


Our results are also consistent with other experiments on \nuc{6}{He} induced reactions on 
several targets, such as  \nuc{64}{Zn} \cite{DiPiet04}, \nuc{65}{Cu} \cite{Nav04} and   \nuc{238}{U} \cite{Raa04}, for which large 1n and 2n transfer cross sections 
have been inferred using different experimental techniques.

In the case of the reaction using the \nuc{238}{U} target, these conclusions 
are supported by the
recent calculations of C\'ardenas {\em et al.}   
\cite{Car06}.
Using 
a schematic coupled-channels calculations, the authors show that the 2n removal cross
section from \nuc{6}{He} can be well accounted for by an incoherent superposition of several
transfer processes with different $Q$ values. Similarly, the  calculations  in this  work involve also
an incoherent sum of 2n transfer to final states with a wide range of
$Q$ values.

\section{\label{sec:summary} Summary and outlook}

In this paper we have presented new experimental data for the reaction
\nuc{6}{He}+\nuc{208}{Pb} at energies around and below the Coulomb barrier. 
The  reaction cross section at backward angles is dominated by a  prominent 
\nuc{4}{He} group, which was interpreted as coming from projectile fragmentation. 

For each scattering energy, the angular and energy distributions of the 
\nuc{4}{He} fragments  have been analyzed and  
compared with  transfer to the continuum calculations, in which the two-neutrons are assumed 
to be transferred as a cluster to both bound as well as unbound states of the target nucleus.  
According to these calculations, most of the observed  $\alpha$ yield comes from the 
transfer of the valence neutrons to  highly excited states of the target in the proximity
of the two-neutron breakup threshold. By contrast, direct breakup calculations and one neutron 
transfer calculations fail to explain the present data.

This analysis 
suggests a scenario in which the \nuc{6}{He} 
nucleus is broken up in the field of the target, and 
the valence neutrons are left with a small relative energy with respect to the
target. By energy conservation, the kinetic energy lost by the neutrons is transferred to 
the $\alpha$ particles, which are therefore accelerated with respect to the beam velocity. 
These
 conclusions are consistent with previous measurements for other reactions induced by 
 \nuc{6}{He} on several targets, for which large transfer cross section have been also reported \cite{Agu00,DeY05,DiPiet04,Nav04,Raa04}.

Further measurements for this reaction, including complete kinematics, wider angular coverage,
detection of fission and evaporation products, etc, would be very useful 
to disentangle more clearly the importance of different reaction mechanisms, and improve 
our understanding of the processes that take place in the scattering of exotic nuclei 
at energies around the Coulomb barrier. Some of these measurements have been  already 
performed and the subsequent analysis is underway.

Despite the reasonable quantitative agreement with the data, the transfer to the continuum  calculations 
can be improved in both the structure and reaction aspects. Concerning the structure model, 
the main approximation involved in our calculation is the assumption of the validity 
of the di-neutron model. Four-body DWBA calculations have been performed by Chatterjee {\em et al.} ~\cite{Chat01} for the 
scattering of \nuc{6}{He} on {Pb} and Au at high energies, using 
a realistic three-body description of the  \nuc{6}{He} nucleus. By comparing with a 
conventional three-body DWBA calculation based upon a di-neutron model of \nuc{6}{He}, they  
find that the latter calculation gives too large breakup cross section. The disagreement is 
attributed to the bigger rms radius in the  di-neutron model, as compared to a realistic 
three-body model. In our calculations this effect is accounted for by increasing the 2n separation 
energy in \nuc{6}{He}. Also, the description of the final states in  \nuc{210}{Pb}
could be improved by using a more realistic level density for this nucleus. 

Concerning 
the description of the reaction mechanism, 
these calculations could be improved by including 
couplings among 
final states in  \nuc{210}{Pb}, by means of the CCBA method. These couplings could be 
generated in a cluster model, by folding the 2n-$\alpha$ and $\alpha$+\nuc{208}{Pb} 
interactions with the internal wavefunctions for the \nuc{210}{Pb} states. 
Also, 
couplings between 
the transfer/breakup channels and the elastic channel could be incorporated beyond the 
first order, thus performing a coupled-reaction channels  calculation. This calculation 
would permit permit an assessment of whether the explicit 
inclusion of these channels can explain the long-range absorption effect encountered 
in the analysis of the elastic data. If this is the case, the optical potential required
to reproduce the elastic data in presence of transfer channels should have a smaller 
diffuseness, as compared to the phenomenological optical potential derived in absence
of these channels. 
Notice that, in the DWBA  approach, these higher order effects are accounted for in an 
effective way by an appropriate choice of the phenomenological optical potentials used to 
describe the incoming and outgoing distorted waves. These CCBA and coupled reaction channels calculations are 
beyond the scope of this work, and hence they have not  been attempted.

%
%
\ack This work has been 
supported by the 
Spanish MCyT projects FPA2005-04460, FPA2005-02379, FPA-2000-1592-C03-02, FPA2003-05958, 
and FPA2002-04181-C04-02/03,  by the European Community-Access to Research Infrastructure action
of the Improving Human Potential Program, contract number HPRI-CT-1999-00110 and the Belgian program
P5/07 on interuniversity attraction poles of the Belgian-state Federal Services for Scientific, Technical and Cultural Affairs.  We are grateful to J. Kolata for useful discussions. A.M.M. acknowledges financial support by the Junta de Andaluc\'{\i}a. D.E. acknowledges financial support by the MEC.


\bibliographystyle{elsart-num}
\bibliography{./he6pb}

\end{document}